\documentclass[aps,pra,superscriptaddress,twocolumn,showpacs]{revtex4}
\usepackage{hyperref}
\usepackage{bm}
\usepackage{epsfig}
\renewcommand*{\[}{\begin{equation}}
\renewcommand*{\]}{\end{equation}}

\begin{document}
\title{A general approach to few-cycle intense laser interactions with complex atoms}
\author{Xiaoxu~Guan,$^1$ O.~Zatsarinny,$^1$ K. Bartschat}
\affiliation{Department of Physics and Astronomy, Drake University, Des Moines, IA 50311, USA}
\author{B.I.~Schneider}
\affiliation{Physics Division, National Science Foundation, Arlington, Virgina 22230, USA}
\author{J.~Feist}
\affiliation{Institute for Theoretical Physics, Vienna University of Technology, A-1040 Vienna, Austria}
\author{C.J.~Noble$^{1,}$}
\affiliation{Computational Science and Engineering Dept., Daresbury Laboratory, Warrington WA4 4AD, UK}
\date{\today}

\begin{abstract}
A general {\it ab-initio\/} and non-perturbative method to solve the time-dependent Schr\"odinger equation (TDSE) for the interaction of 
a strong attosecond laser pulse with a general atom, i.e., beyond the models of quasi-one-electron or quasi-two-electron targets, 
is described. 
The field-free Hamiltonian and the dipole matrices are generated using a flexible $B$-spline $R$-matrix method.
This numerical implementation enables us to construct term-dependent, 
non-orthogonal sets of one-electron orbitals for the bound
and continuum electrons. The solution of the TDSE is propagated in time using the 
Arnoldi-Lanczos method, which does not require the diagonalization of any large matrices.  
The method is illustrated by an application to the multi-photon excitation and ionization of Ne atoms.
Good agreement with $R$-matrix Floquet calculations for the generalized cross sections for two-photon
ionization is achieved.
\end{abstract}

\pacs{32.80.Fb,32.80.Rm,42.65.Re}
\maketitle

\section{Introduction}
The ongoing development of ultra-short and ultra-intense light sources based on
high-harmonic generation and free-electron lasers is providing new ways to generate optical pulses capable
of probing dynamical processes that occur on atto\-second time scales~\cite{KITP_AS_Workshop}. 
These atto\-second pulses are providing a window to study the details of electron interactions
in atoms and molecules in the same way that femto\-second pulses revolutionized 
the study of chemical processes.  Single atto\-second pulses or pulse trains open
up new avenues for time-domain studies of multi-electron dynamics in atoms, molecules, 
plasmas, and solids on their natural, quantum mechanical time scale and at distances 
shorter than molecular and even atomic dimensions. These capabilities promise a revolution
in our microscopic knowledge and understanding of matter~\cite{Krausz}.  A major role for theory in 
atto\-second science is to elucidate novel ways to investigate and to control electronic 
and other processes in matter on such ultra-short time scales. 

The ingredients of an appropriate theoretical and computational formulation require an accurate and efficient generation
of the Hamiltonian and electron$-$field interaction matrix elements, as well as 
an optimal approach to propagate the time-dependent Schr\"odinger equation (TDSE).  
Many theoretical papers have been devoted to the propagation of the TDSE including laser pulses.  
The earliest calculations employed finite-difference methods~\cite{KuKrSc} to discretize the spatial 
coordinates.  As shown in a recent review by Pindzola {\it et al}.~\cite{Pind07}, this 
method is still being used with great success today.  Other formulations
employ finite-element~\cite{Bottcher}, discrete-variable, or finite-element discrete-variable representation (FEDVR) 
\cite{Light-Hamilton-Lill,Rescigno-McCurdy,Schneider-Collins-Hu} 
approaches to discretize the coordinates and thereby take 
advantage of the higher accuracy afforded by these methods. Time propagation of the 
wavefunction may also be accomplished by a variety
of techniques.  These include simple approaches such as the leapfrog or 
Runge-Kutta~\cite{Numerical Recipes} method to more sophisticated split-operator~\cite{split} or Krylov space 
iterations~\cite{Park-Light,Krylov}.    
A selected set of references is given in the bibliography.  The relevant physical 
information is extracted from the TDSE by projecting the wavefunction 
onto appropriate long-range solutions after the laser interaction has vanished.  The 
details of the process depend on what parameters are desired; total ionic yields are 
relatively simple to extract while differential or doubly differential quantities 
necessitate more work~\cite{Pind07}. 
\par
In this paper we consider a new approach to model the interaction of an atomic
system with a strong laser pulse. We combine a highly flexible \hbox{$R$-matrix} method~\cite{ZFF00,ZB04,Zat06},
including non-orthogonal sets of atomic orbitals to describe the initial bound state as 
well as the ejected-electron$-$residual-ion interaction, with the 
Arnoldi-Lanczos iterative propagation scheme.  In contrast to many other methods currently being used for such problems 
\cite{Hu-Collins-1,Hu-Collins-2,KTaylor-et-al}, the present implementation is not 
restricted to (quasi-)one- or (quasi-)two-electron targets.  It can be applied to 
{\it complex\/} atoms, such as inert gases other than helium and even open-shell systems 
with non-vanishing spin and orbital angular momenta.
We illustrate the method with results for multi-photon excitation and ionization of 
neon by a linearly polarized laser pulse.

\section{Numerical Method}
\subsection{The B-Spline R-Matrix  Method}
Unless specified otherwise, atomic units are used throughout this manuscript.
The TDSE for the $N$-electron wavefunction $\Psi(\bm{r}_1,\ldots,\bm{r}_N;t)$ of the 
present problem is given by
\begin{eqnarray}
\label{TDSE}
i {\partial \over \partial t} \Psi(\bm{r}_1,...,\bm{r}_N;t) &=& \nonumber \\
  \big[ \bm{H}_0(\bm{r}_1,...,\bm{r}_N) \!\!&+&\!\! V(\bm{r}_1,...,\bm{r}_N;t) \big] \Psi(\bm{r}_1,...,\bm{r}_N;t),
\nonumber \\
\end{eqnarray}
where $\bm{H}_0(\bm{r}_1,...,\bm{r}_N)$ is the field-free Hamiltonian containing the sum of the kinetic energy of the $N$ 
electrons, their potential energy in the field of the nucleus (we assume an infinite nuclear mass), and
their mutual Coulomb repulsion, while $V(\bm{r}_1,...,\bm{r}_N;t)$ represents the interaction of the electrons with the 
electromagnetic field.  

We expand the wavefunction as
\begin{equation} 
\label{expand}
\Psi(\bm{r}_1,...,\bm{r}_N;t) = \sum_q C_q(t) \Phi_q(\bm{r}_1,...,\bm{r}_N).
\end{equation}
Here $\Phi_q(\bm{r}_1,...,\bm{r}_N)$ are a known
set of $N$-electron states formed from appropriately symmetrized products of 
atomic orbitals. Optimization procedures tailored to the individual neutral, ionic, and continuum orbitals may be employed 
since the atomic one-electron orbitals are not forced to be orthogonal.  The radial parts of the atomic orbitals are themselves 
expanded in $B$-splines.  In the practical implementation of the $B$-spline $R$-matrix (BSR) method~\cite{Zat06}, factors 
that depend on angular and spin momenta are separated
from the radial degrees of freedom.  This enables the production of a ``formula tape'' since
many Hamiltonian matrix elements share common features.  Given a set of atomic orbitals, it
is possible to realize a great economy in the construction of the actual Hamiltonian matrix using this symbolic tape 
even for non-orthogonal basis sets, since ultimately every matrix element is a linear combination of
one-electron and two-electron radial integrals multiplied by overlaps and angular factors.

A significant advantage of the BSR method in the calculation of both bound and continuum
states is the possibility of employing non-orthogonal sets of atomic orbitals for different target states, 
thereby omitting the need for pseudo-orbitals to account for the strong term-dependence that  
exists in many complex targets, with the noble gases being a prime example.  Furthermore, we do not force
the partial wave describing a continuum electron with orbital angular momentum~$\ell$ to be orthogonal to all 
bound orbitals with the same~$\ell$.  While the method gives great flexibility in the target description,
allowing for accurate representations with relatively small configuration interaction expansions, and
also simplifies the general form of the close-coupling expansion used to generate the 
bound and excited states of the atomic system, the price to pay is the representation of
the field-free Hamiltonian and the dipole matrices in a non-orthogonal basis. If desired, the non-orthogonality
of the primitive $B$-spline basis could easily be removed by replacing the splines  by another
complete but orthogonal basis, e.g., a finite-element discrete-variable representation~\cite{Schneider-Collins-Hu}.
However, if one wants the flexibility associated with a non-orthogonal set of physical orbitals expanded in any 
primitive basis, it is necessary to deal with the non-orthogonality issues directly.
\par
The interaction of the atomic electrons with the time-dependent
electric potential, in the length form of the electric dipole approximation, is given by
\begin{equation}
\label{length}
V(\bm{r}_1,\ldots,\bm{r}_N;t) = \sum_{i=1}^N \bm{E}(t) \cdot  \bm{r}_i
\end{equation}
where $\bm{E}(t)$ is the electric field.  This form has been used for the calculations in this paper. 
For simplicity of the notation, we have omitted the spin-coordinates of the electrons.  Since the initial bound state 
is a singlet state in our case, only singlet states will have to be coupled in the subsequent partial-wave
expansion. 

The tasks at hand are now i)~the preparation of the initial state, ii)~the time propagation
of the $C_q(t)$, 
and iii)~the extraction of physically relevant information from the final state after 
the time propagation.  As mentioned above, the present approach employs the BSR method 
described in 
refs.~\cite{ZFF00,ZB04,Zat06} 
to compute all the time-independent matrix elements needed for the problem.
These parts require a representation of the field-free Hamiltonian matrices for the 
partial-wave symmetries \hbox{$^1S^e$}, \hbox{$^1P^o$}, \hbox{$^1P^e$}, \hbox{$^1D^e$}, \hbox{$^1D^o, \ldots$}, as well as the
dipole matrices that couple any given value of the total orbital angular momentum~$L$ with a 
given parity to the symmetries with $L$ and $L\pm 1$ of the opposite parity. 
All of these matrices can be readily generated with the BSR method, which may also be used to represent the initial bound state.  
Since the time dependence of the Hamiltonian 
appears as a simple multiplicative factor, this only needs to be done {\it once\/} at the beginning
of the calculation.
When the expansion in (\ref{expand}) is inserted into the Schr\"odinger equation, we obtain,
\begin{equation}
\label{matrix form}
i \bm{S} {\partial \over \partial t} \bm{C} = \big[ \bm{H}_0(\bm{r}_1,...,\bm{r}_N) \!+\! \sum_{i=1}^N \bm{E}(t) \cdot  \bm{r}_i \big]  \bm{C},
\end{equation}
where $\bm{S}$ is the overlap matrix of the basis functions.
Since we are initially interested in excitation and single ionization of the target atom by the laser pulse, the
symmetries of the field-free Hamiltonian must also contain a sufficient number of singly excited bound
states as well as the continuum states representing electron scattering from the residual ion. As a method
developed to treat exactly such problems, the BSR approach is particularly suitable to represent these states.

\subsection{Time Propagation}
Time propagation of the initial wavefunction may be accomplished using a number of approaches.  
Explicit, norm-conserving approaches, which rely on simple matrix-vector multiplication, 
are generally preferred to implicit methods, which require the solution of a set of linear equations.  
Of the former methods, we found the Arnoldi-Lanczos approach~\cite{Park-Light,Schneider-Collins} 
to be quite effective, provided one is able to deal with the (often) poorly conditioned matrices generated by 
non-orthogonal basis sets. A general discussion and error analysis of the Arnoldi-Lanczos method can be
found in the work of Saad~\cite{Saad}. Here we only sketch the basic ideas
relevant to our solution of the TDSE. 

A straightforward approach is to transform the non-orthogonal many-electron basis to an orthogonal basis using 
the L\"owdin transformation to generate new field-free Hamiltonian 
and dipole matrix blocks through
\begin{eqnarray}
\bm{H}_0' &=& \bm{S}^{-1/2} \bm{H}_0 \bm{S}^{-1/2}, \\
\bm{D}' &=& \bm{S}^{-1/2} \bm{D} \bm{S}^{-1/2}.
\end{eqnarray}
We thus have 
\begin{equation}
i\frac{\partial}{\partial t} \bm{C}' = \bm{H}'(t)\bm{C}' ,
\end{equation}
where $\bm{H}'(t)= \bm{H}_0' + E(t) \bm{D}'$.
Since $\bm{H}_0$, $\bm{D}$, and $\bm{S}$ are all time-independent, this only requires the diagonalization of the 
overlap matrix {\it once}, and a few matrix-vector multiplications at every time step. 

The essential idea of the Arnoldi-Lanczos method is to construct a reduced Krylov space of dimension $m$,  at time~$t + \Delta t$, 
\begin{equation}
{\cal K}_m(\bm{H}',\bm{v}) \equiv {\rm span} \{ \bm{v}, \bm{H}'\bm{v}, \bm{H}'^2\bm{v}, 
         \ldots,\bm{H}'^{(m-1)}\bm{v} \},
\end{equation}
where the initial  vector~$\bm{v}$ is the previously computed solution at time~$t$.  These vectors, which are generated by repeatedly applying the Hamiltonian $\bm{H}'(t)$ on the vector~$\bm{v}$, are not used directly, but orthonormalized using the Lanczos recursion,
\begin{equation}
\label{Lanczos recursion}
\beta_{n+1} \bm{ v}_{n+1} = ( \bm{H}' - \alpha_n ) \bm{v}_n - \beta_n \bm{v}_{n-1}
\end{equation}
 to transform the Hamiltonian matrix to tridiagonal form~\cite{Saad} as long as the original matrix is Hermitian.   The elements, $\alpha_n$ and $\beta_n$, of the tridiagonal matrix, may be computed (see below for a slightly more general case) during the recursion process using simple scalar products.
The resultant tridiagonal matrix is then diagonalized using standard algorithms.    The result of the above procedure is an $N \times m$ matrix $\bm{Q}$, which transforms the matrices from $\bm{H}'$ with rank~$N$ to $\bm{h}$ with rank~$m$.  Finally, the time evolution from $t$ to $t+\Delta t$ is achieved through 
\begin{equation}
\bm{C}'(t+\Delta t) = \bm{Q} e^{-i \bm{h} \Delta t} \bm{Q}^\dagger \bm{C}'(t).
\end{equation}
At each step $m$ of the process, a convergence test is performed and once the propagated solution from two successive time steps has fallen below a predetermined criterion, the recursion is terminated.
As long as the rank~$m$  of the process is substantially smaller than the original matrix size $N$, the process can be very effective.  Finally, we note that the Arnoldi-Lanczos algorithm outlined above conserves the norm, i.e.,
$|\bm{C}'(t+\Delta t)|^2 = |\bm{C}'(t)|^2$.
 
One of the appealing features in the Arnoldi-Lanczos
procedure is the fact that only matrix-vector multiplications and scalar products are required. This allows us 
to take advantage of specific algorithms if the matrix is sparse. 
It is also worthwhile to make a few remarks regarding the size of the Krylov space and the
time step to obtain stable and accurate solutions.  Since we need to generate new vectors in the
Krylov space repeatedly and hence want to keep the size of
that space manageable in practical calculations, we can only take
relatively small time steps. For
most calculations presented in this paper, 1000 time steps per optical cycle were used
to propagate the system.  Not surprisingly, as already noted by Park and Light~\cite{Park-Light}, 
numerical experiments showed that enlarging the Krylov space allows for larger time steps to be
taken.  This relationship was used to optimize the efficiency of our algorithm. 

An alternative, theoretically equivalent approach generalizes the Lanczos process to a non-orthogonal basis.
It thus allows for directly solving Eq.~(\ref{matrix form}) without transforming it to an orthogonal basis.  
In this case, we use the recursion
\begin{equation}
\label{recursion}
\beta_{n+1} \bm{S v}_{n+1} = ( \bm{H} - \alpha_n \bm{S} ) \bm{v}_n - \beta_n \bm{S v}_{n-1} = \bm{q}_n,
\end{equation} 
where the $\bm{v}_n$ are the Lanczos vectors.  These vectors are required to satisfy the condition,
\begin{equation}
 \bm{v}_n^{\dagger} \bm{S v}_m = \delta_{n,m}.
\end{equation}
This so-called $S$-orthogonalization is possible since $\bm{S}$ is positive definite.  The calculation proceeds
along the following steps.  After computing
\begin{equation}
\label{alpha}
\alpha_n = \bm{v}_n^{\dagger} \bm{H} \bm{v}_n,
\end{equation}
the $\bm{q_n}$ may be generated through matrix-vector multiplication and previously obtained coefficients.  
The next step computes
\begin{equation} 
\label{beta}
\beta_{n+1} = \sqrt { \bm{q}_{n}^{\dagger} \bm{S}^{-1} \bm{q}_n }.
\end{equation}
In practice, no matrix inversions are performed.  Instead, the $\bm{S}$-matrix is decomposed using the 
Cholesky decomposition for positive definite matrices,
\begin{equation} 
\bm{S} = \bm{L^{ \dagger } L},
\end{equation}
at the beginning of the calculation. This yields
\begin{equation} 
\label{final beta}
\beta_{n+1} = \sqrt { \bm{T}_n^{\dagger} \bm{T}_n } ,
\end{equation}
with
\begin{equation} 
\bm{T}_n = (\bm{L}^{-1})^\dagger \bm{q}_n ,
\end{equation}
and only requires the solution of a triangular set of linear equations once the Cholesky decomposition is performed.  
To complete the calculation it is necessary to solve a second set of triangular equations, namely
\begin{equation} 
\bm{v}_{n+1} = \bm{L}^{-1} \bm{T}_{n} / \beta_{n+1}.
\end{equation}
From a numerical point of view, the Cholesky decomposition is somewhat cheaper than the diagonalization of the overlap matrix, while
the solution of the triangular sets of linear systems at each iteration is comparable in cost to the matrix-vector 
multiplication.  For the case of an orthonormal basis, $\bm{S=I}$, the algorithms are identical.  
More importantly for future work, we note that there are other 
possibilities, which entirely avoid the need for either inverting or decomposing any large matrices during the calculation~\cite{Golub-Yee}.

For the present work, we implemented the Arnoldi-Lanczos method with a fixed size of the Krylov space. 
As expected, results obtained with either of the above methods were identical, and the matrix sizes we had to deal with
(ranks of less than 500 for each individual block of the field-free Hamiltonian and the dipole matrix) 
were so small that we could actually check the results by performing an exact diagonalization.
A well-known alternative regarding the size of the Krylov space
requires checking the convergence at each step and, if necessary, augmenting the size of the space.  
By comparing results obtained with different sizes of the Krylov space, we ensure numerical convergence of the final
results with the size of that space.  

In addition, we check the dependence of the final results on the
number of coupled symmetries.  It is well known~\cite{van der Hart} that the number of $L$-values to couple increases
strongly with decreasing laser frequency. Employing the velocity gauge to express the dipole operator is expected to 
reduce this problem~\cite{CL96}. However, the velocity gauge is problematic at short distances, and the problems increase with
the nuclear charge of the target~\cite{vdH2007}. Hence, we plan to explore switching between gauges at some distance 
in future work.
Finally, to ensure that box effects do not disguise the actual physics, we also
use a masking function to avoid reflection from the boundaries of our box.

\section{Application: Multi-photon Ionization of Ne}
As a first application of our approach, we studied the short-pulse, multi-photon ionization of Ne.
The electric field was taken as linearly polarized.  
We only accounted for single-electron excitation and 
ionization leading to $1s^2 2s^2 2p^5 n\ell$ bound states or $1s^2 2s^2 2p^5 k\ell$ 
continuum states in the present, proof-of-principle, calculation.  Extensions to handle 
more complex excitations and/or double ionization are possible and will be discussed in 
the conclusion of the paper.
  
After the wavefunction has been propagated,
the relevant information is extracted by standard projection techniques.  This requires
the ground-state wavefunction of the Ne atom, $\Psi_0$, obtained either by imaginary time 
propagation or exact diagonalization, and the unperturbed states, $\Psi^0_{\gamma,L}$, where
the label $\gamma$ represents the collection of quantum numbers needed to define the 
state of a bound or free electron in the field of the residual atomic ion,
{\it asymptotically}.  In practice, these states are constructed from a linear combination 
of products of bound excited states of the Ne$^+$ ion coupled to a bound or continuum
function of the additional electron.  

The quantities of interest for the present work are the total survival probability of the initial state,
\begin{equation}
\label{Total}
P_{0} = |< \Psi_0 | \Psi >|^2,
\end{equation}
 the probability of finding a given ($\gamma, L$) state,
\begin{equation}
\label{fractional}
P_{\gamma,L} = |< \Psi^{0}_{\gamma,L} | \Psi >|^2,
\end{equation}
and the total probability into a specific $L$,
\begin{equation}
\label{fractional L}
P_{L} = \sum_{\gamma}  P_{\gamma,L}.
\end{equation}
In computing $P_{\gamma,L}$, the time-propagated wavefunction is projected onto the singly excited states with an energy 
below (excitation) or above (ionization) the single-ionization 
threshold leading to Ne$^+$.  In practice, we compute the bound-state fraction and get the contribution from the continuum by 
subtraction plus the loss in the norm due to the masking function.  
\begin{figure}[tbp]
\includegraphics[width=0.98\columnwidth]{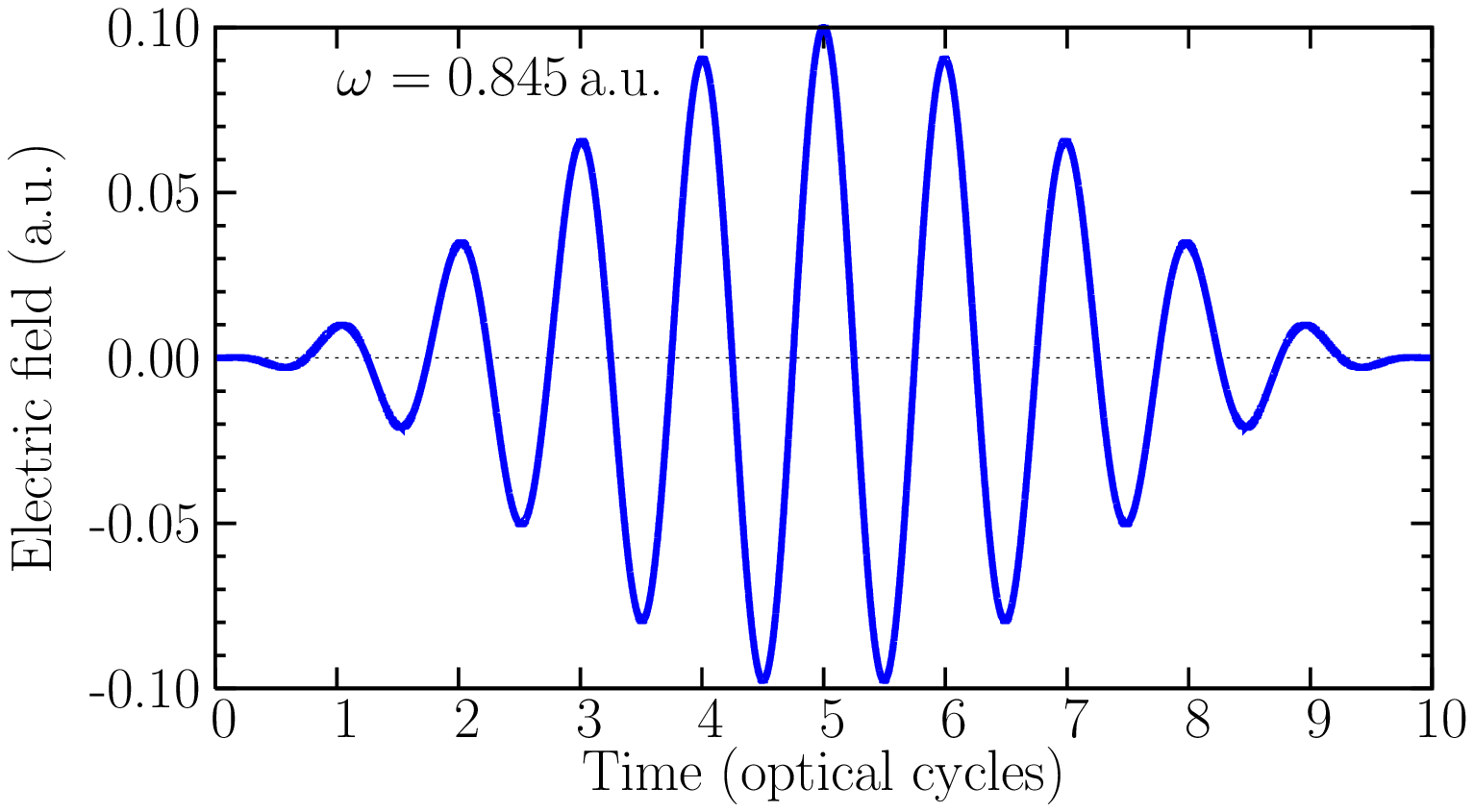}
\par
\includegraphics[width=0.98\columnwidth]{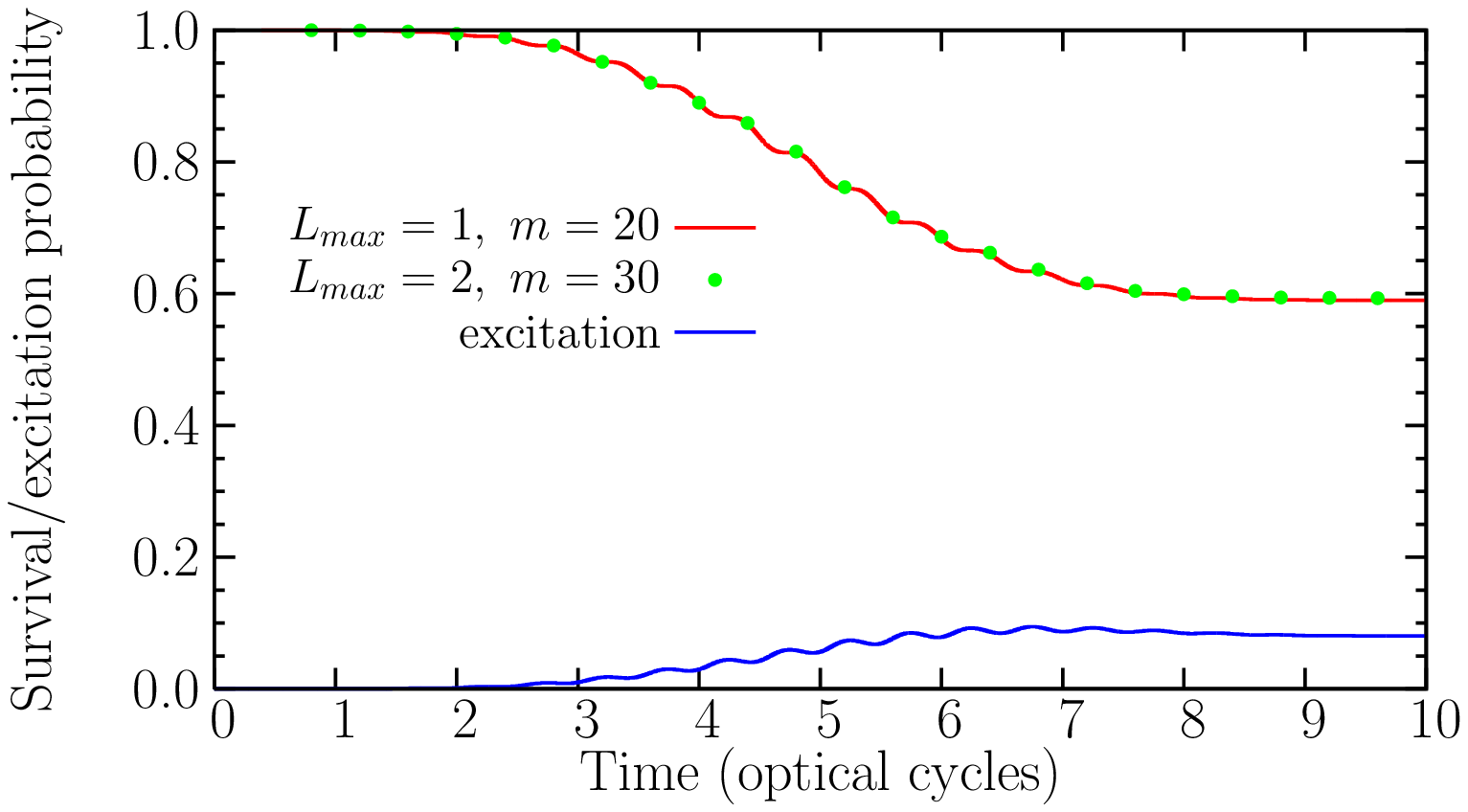}
\par
\includegraphics[width=0.98\columnwidth]{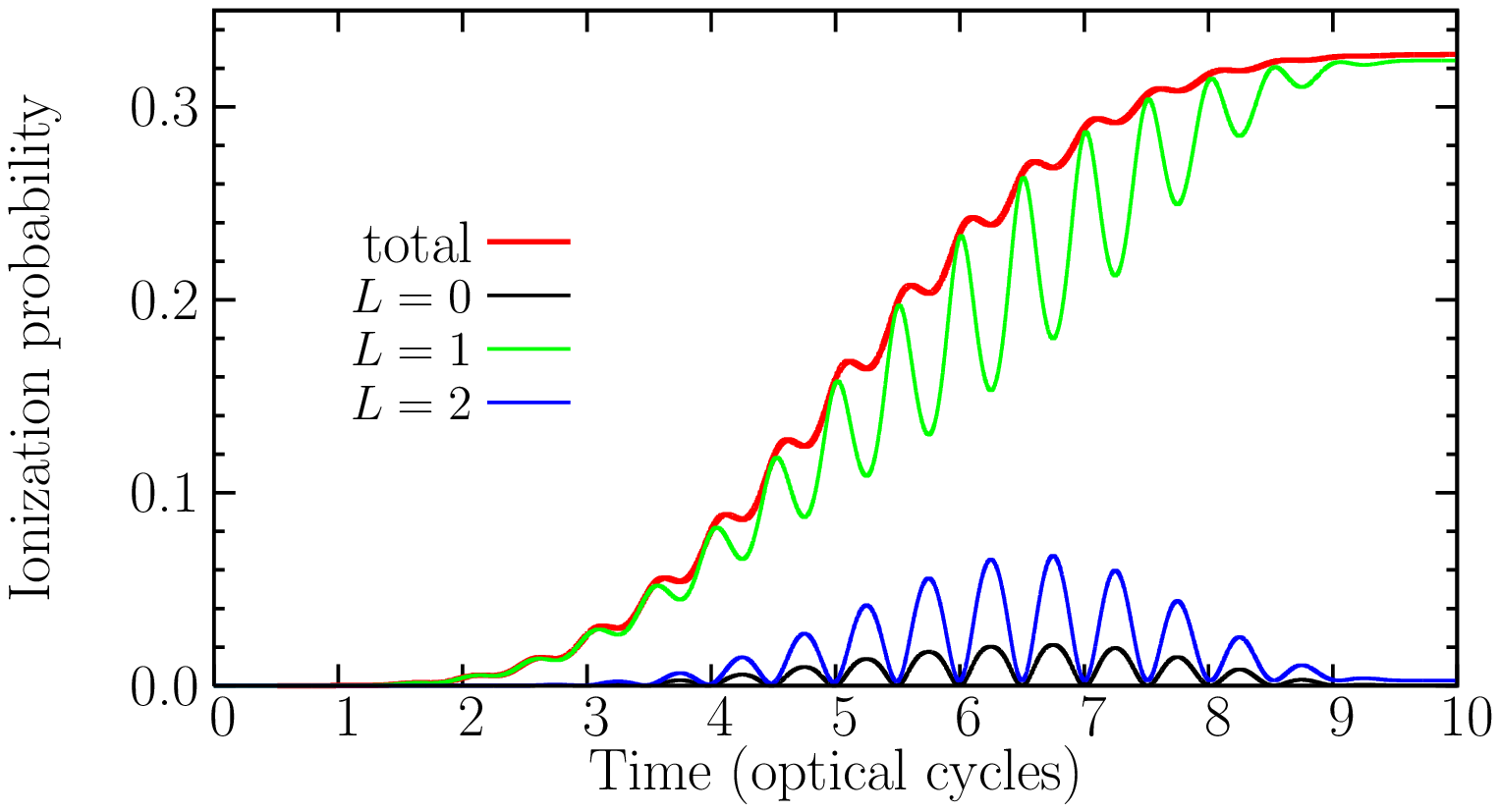}
\caption{(Color online). Laser pulse (top panel), ground-state survival and excitation probabilities (center), and single-ionization
probabilities (bottom) for the interaction of a short laser pulse of frequency 0.845~a.u.~and 
peak intensity of $\rm 3.5 \times 10^{14}\,W/cm^2$
with a neon atom in its $(2p^6)^1S$ ground state. Using different sizes~$m$ of the Krylov space
and numbers of coupled $L$-values ($L_{max}$), we demonstrate in the center panel the numerical 
convergence of our results for
the survival probability of the ground state as a function of time.  In the bottom panel, we show the contributions 
to the total ionization probability from 
ionization continua with different values of the total orbital angular momentum~$L$.}
\end{figure}

Figure~1 shows our results for the response of a neon atom in its ground state
$(2p^6)^1S$ to the effect of a 10-cycle laser pulse with a sin$^2$ envelope.  Since the
frequency of the laser pulse is large (sufficient to ionize the atom by absorption of a single photon), 
only a few values for the total angular momentum~$L$ of
the system have to be coupled to obtain converged results for single ionization, with relatively
small dimensions~$m$ of the Krylov space.  
As seen from the figure, the ionization process with the highest probability indeed is ionization to the \hbox{$P$-continuum}, i.e.,
effectively a one-photon absorption process leading to a free electron with $\ell = 0$ or $\ell = 2$, respectively.  The 
survival probability for the ground state is approximately~60\%, while the probability 
for excitation is just under~10\%.  Finally, the probability for
the ejected electron to have an orbital angular momentum of~$\ell = 1$ or $\ell = 3$, i.e., forming an
$L=0$ or $L=2$ state of the $\rm e-Ne^+$ scattering problem, is small but not zero.

\begin{figure}[tbp]
\includegraphics[width=0.98\columnwidth]{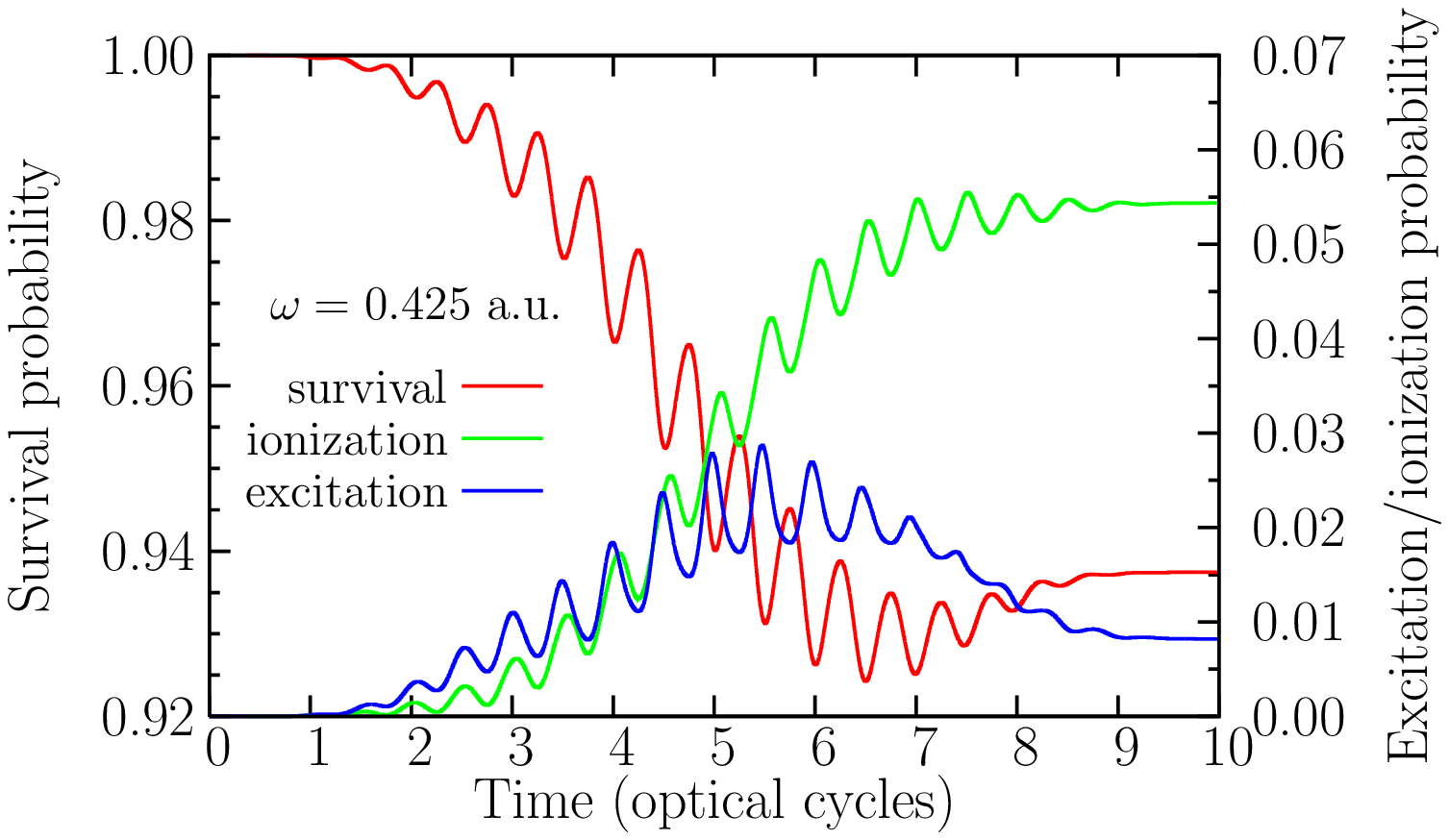}
\par
\includegraphics[width=0.98\columnwidth]{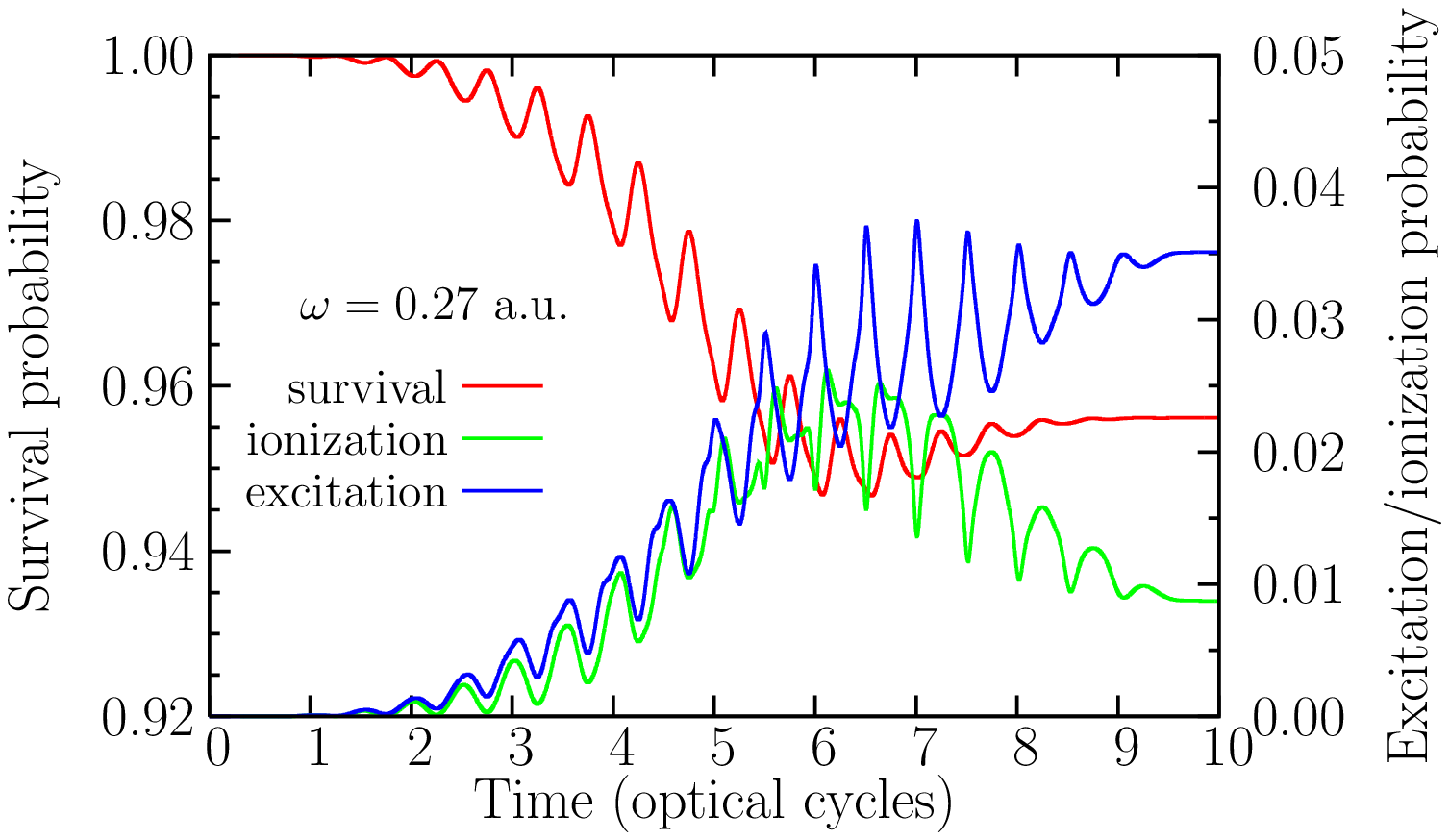}
\caption{(Color online). Ground-state survival (left scale) and total excitation and ionization probabilities (right scale) for
laser frequencies of 0.425~a.u.~and 0.270~a.u.~and a 
peak intensity of $\rm 3.5 \times 10^{14}\,W/cm^2$.}
\end{figure}

Figure~2 shows the response of the Ne atom to pulses with approximately one-half and one-third of the laser frequency used in figure~1.  In this
case, at least two or three photons, respectively, need to be absorbed in order to ionize the system. A significantly larger number of
symmetries (we used $L_{max} = 6$) must be coupled to get converged results for these cases.  Note that excitation rather than 
ionization appears as the dominating reaction process
for $\omega = 0.27\,$a.u. and the laser parameters chosen here. 

The dependence of the ionization probability on the laser intensity is plotted in figure~3 for 
laser frequencies of 0.425~a.u. and 0.270~a.u., respectively. For peak intensities in the range
$\rm 10^{13} - 5 \times 10^{14}\,W/cm^2$, the slopes in the log-log plot (increases of about two or three orders of magnitude in the
probability per one order of magnitude increase in the intensity) are consistent with the expectation from 
lowest-order perturbation theory that ionization is effectively caused by two-photon or three-photon processes, without
hitting any resonances. For higher
laser intensities, the curve flattens because of both saturation and 
double ionization. The description of the latter processes is,
in principle, also possible with the current method.  However, it requires the inclusion of double-continuum states
with a Ne$^{2+}$ core in the current expansion and, therefore, significantly more computational resources. 

\begin{figure}[tbp]
\includegraphics[width=0.98\columnwidth]{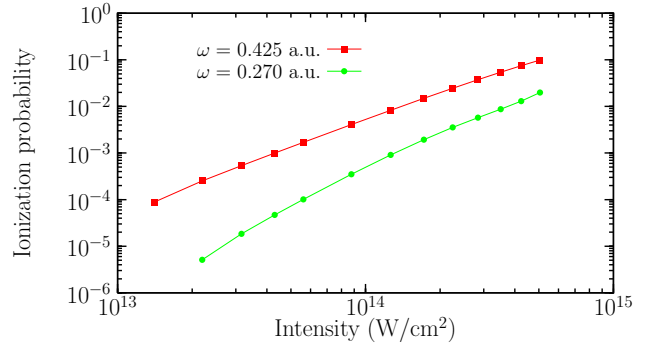}
\caption{(Color online). Ionization probability vs. laser peak intensity at laser frequencies of 0.270~a.u. and 0.425~a.u.}
\end{figure}

\begin{figure}[tbp]
\includegraphics[width=0.98\columnwidth]{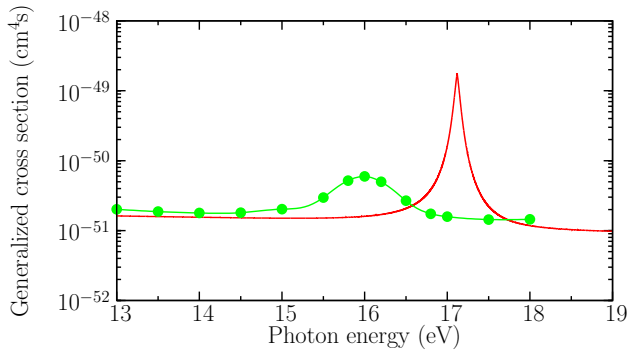}
\caption{(Color online). Generalized cross section for two-photon ionization of 
Ne($2p^6)^1S$ as a function of photon energy.  The current results (circles) are compared
with the $R$-matrix Floquet predictions of McKenna and van der Hart~\cite{McK04}.}
\end{figure}
As a further check of our present work, we now consider the generalized cross section for
two-photon ionization. 
Having obtained the total two-photon ionization rate $\Gamma^{(2)}$ by propagating the wave\-function in 
a longer pulse (30~optical cycles in the present case), the generalized
cross section $\sigma^{(2)}$ for two-photon ionization is obtained as~described by Charalambidis {\it et al}.~\cite{Cha97}.
In figure~4, we compare our
non-perturbative results for a few photon energies to the $R$-matrix
Floquet predictions of McKenna and van der Hart~\cite{McK04}. We note satisfactory
agreement for energies away from the first resonance structure
corresponding to the intermediate $(2p^5 3s)^1P^o$ state. 
Since the Floquet approach effectively 
corresponds to an infinitely long ``pulse'' and hence a sharp photon energy,
it can resolve this structure, while we get a broader maximum due to the
frequency width of our pulse.  The shift in the energy position of the resonance is
due to the different structure models used in the two calculations.

\begin{figure}[tbp]
\includegraphics[width=0.98\columnwidth]{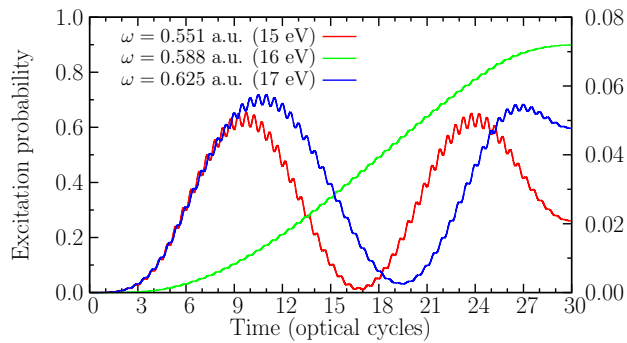}
\caption{(Color online). Excitation probability of the Ne$(2p^5 3s)^1P$ state for a 30-cycle
laser pulse with photon energies of 15~eV (right scale), 16~eV (left scale), and 17~eV (right scale) and a 
peak intensity of $\rm 2.0 \times 10^{13}\,W/cm^2$.}
\end{figure}

To further illustrate the effect of
the resonant $(2p^5 3s)^1P$ state (around 16~eV in our model) 
we present the excitation probability for three photon energies in figure~5.  Away from the
resonance, at 15~eV and 17~eV, there are several Rabi oscillations between the ground state
and the excited state during a 30-cycle pulse, and the maximum probability for excitation is about 5\% during
these oscillations (right scale of figure~5).  On the other hand, we just reach the first maximum in the 
excitation probability for the resonance energy of 16~eV after 30~cycles, and the value of that maximum
is above 90\% (left scale of figure~5). This shows the strong effect of
the energy detuning on the frequency and the amplitude of the Rabi oscillations.

\section{Conclusions and Outlook}
We have described a general method for treating the interaction of 
a strong atto\-second laser pulse with a complex atom.
The approach combines a highly flexible \hbox{$B$-spline} 
\hbox{$R$-matrix} method for the description of the initial state, other bound states in the system, 
the ionic core, and the interaction of the free electron with the residual ion after ionization, 
with an efficient Arnoldi-Lanczos scheme for the time propagation of the TDSE. The major advantages of the method
are i)~its generality and applicability to any complex many-electron target,  ii)~the possibility of 
generating highly accurate target and continuum descriptions with relatively small configuration interaction 
expansions, and  iii)~an efficient time-propagation technique.  In the current paper we limited 
the continuum states to include only singly ionized states.  To extend this to doubly ionized targets requires 
that we allow two \hbox{$R$-matrix} orbitals outside a doubly charged ionic core. In principle, this is a 
straightforward extension of the current codes, but in practice the size of the matrix blocks will increase 
dramatically.  The critical advantage of the present approach is that non-orthogonal basis sets 
should significantly reduce the size of the configuration expansion compared to approaches based on orthogonal sets. 
 
In the future, we plan to further analyze and improve the numerical efficiency of the method, particularly by investigating
different schemes of setting up the matrices. Prime candidates are expansions in other complete bases such as 
finite-element discrete-variable representations.  The use of many-electron expansions in non-orthogonal basis sets  
also necessitates developing efficient, new approaches to the time propagation of the wavefunction.  While we have described one 
possibility in this paper, it is not the only one and likely far from the best approach.  We are actively investigating 
other methods, using approximate and easily computed inverses,  which do not require any factorization or 
diagonalization of large matrices.  Most critically, we need to extend the current BSR method to treat two 
free electrons outside an ionic core, if we are to treat problems involving multi-photon double-ionization 
of complex targets and compare with recent free-electron laser experiments such as those reported in refs.~\cite{Sorokin,Moshammer}.
Finally, looking at angle-differential observables will also require a reliable method to extract the relevant
information, such as amplitudes from single and double ionization, from the time-propagated wavefunction~\cite{McCurdy04}.
All of these issues are currently under active investigation by our collaboration.

\section*{Acknowledgments}
The authors would like to thank Hugo van der Hart for helpful discussions and for making results available in electronic form.
This work was supported by the United States National Science Foundation
under Grants No. PHY-0244470~(XG,KB) and PHY-0555226 (OZ,KB,CJN).

\end{document}